\documentclass[
twocolumn]{aastex631}
\usepackage{graphicx}%
\usepackage{multirow}%
\usepackage{amsmath,amssymb,amsfonts}%
\usepackage{amsthm}%
\usepackage{mathrsfs}%
\usepackage[title]{appendix}%
\usepackage{xcolor}%
\usepackage{textcomp}%
\usepackage{booktabs}%
\usepackage{algorithm}%
\usepackage{algorithmicx}%
\usepackage{algpseudocode}%
\usepackage{listings}%
\usepackage[normalem]{ulem}%
\usepackage{amsmath}
\usepackage{soul}

\newcommand{\dM}{\textit{dM}}
\newcommand{\dMe}{\textit{dMe}}
\newcommand{\ddcr}{\ensuremath{\Delta\mathrm{DCR}}}
\newcommand{\eg}{\textit{e.g.}}
\newcommand{\ie}{\textit{i.e.}}
\newcommand{\new}[1]{{\color{black} #1}}

\theoremstyle{thmstyletwo}%

\theoremstyle{thmstylethree}%

\begin{document}

\title{First Temperature Profile of a Stellar Flare using Differential Chromatic Refraction}

\author[0000-0001-9273-5036]{Riley W. Clarke}
\affiliation{Department of Physics \& Astronomy,
University of Delaware,
Newark, DE 19716, USA}
\affiliation{Data Science Institute,
University of Delaware,
Newark, DE 19716, USA}

\author[0000-0003-1953-8727]{Federica Bianco}
\affiliation{Department of Physics \& Astronomy,
University of Delaware,
Newark, DE 19716, USA}
\affiliation{Joseph R. Biden School of Public Policy and Administration,
University of Delaware,
Newark, DE 19716, USA}
\affiliation{Data Science Institute,
University of Delaware,
Newark, DE 19716, USA}

\author[0000-0002-0637-835X]{James R. A. Davenport}
\affiliation{DiRAC Institute, Department of Astronomy,
University of Washington,
Seattle, WA 98195, USA}

\author[0000-0001-5703-2108]{Jeffery Cooke}
\affiliation{Centre for Astrophysics and Supercomputing, Swinburne University of Technology, John St, Hawthorn, VIC 3122, Australia}
\affiliation{ARC Centre of Excellence for Gravitational Wave Discovery (OzGrav), Hawthorn, 3122, Australia}

\author[0000-0003-2601-1472]{Sara Webb}
\affiliation{Centre for Astrophysics and Supercomputing, Swinburne University of Technology, John St, Hawthorn, VIC 3122, Australia}
\affiliation{ARC Centre of Excellence for Gravitational Wave Discovery (OzGrav), Hawthorn, 3122, Australia}

\author[0000-0002-8977-1498]{Igor Andreoni}
\affiliation{University of North Carolina at Chapel Hill, 120 E. Cameron Ave., Chapel Hill, NC 27514, USA}

\author[0000-0001-9227-8349]{Tyler Pritchard}
\affiliation{NASA’s Goddard Space Flight Center, Greenbelt, MD 20771 USA}

\author[0000-0001-5326-3486]{Aaron Roodman}
\affiliation{Kavli Institute for Particle Astrophysics and Cosmology, PO Box 2450,
Stanford University, Stanford, CA 94305, USA}
\affiliation{SLAC National Accelerator Laboratory, Menlo Park, CA 94025, USA}

\begin{abstract}
We present the first derivation of a stellar flare temperature profile from single-band photometry. Stellar flare DWF030225.574-545707.45129 was detected in 2015 by the Dark Energy Camera as part of the \textit{Deeper, Wider, Faster} Programme. The brightness ($\Delta m_g = -6.12$) of this flare, combined with the high air mass ($1.45 \lesssim X \lesssim 1.75$) and blue filter (DES $g$, 398-548 nm) in which it was observed, provided ideal conditions to measure the zenith-ward apparent motion of the source due to differential chromatic refraction (DCR) and, from that, infer the effective temperature of the event. We model the flare's spectral energy distribution as a blackbody to produce the constraints on flare temperature and geometric properties derived from single-band photometry. We additionally demonstrate how simplistic assumptions on the flaring spectrum, as well as on the evolution of flare geometry, can result in solutions that overestimate effective temperature. Exploiting DCR enables studying chromatic phenomena with ground-based astrophysical surveys and stellar flares on M-dwarfs are a particularly enticing target for such studies due to their ubiquity across the sky, and the heightened color contrast between their red quiescent photospheres and the blue flare emission. Our novel method will enable similar temperature constraints for large sample of objects in upcoming photometric surveys like the Vera C. Rubin Legacy Survey of Space and Time.
\end{abstract}

\section{Introduction} \label{sec:intro}


Differential Chromatic Refraction (DCR, \citealt{filippenko1982}) is an atmospheric effect that causes astrophysical sources to fall on different locations of the focal plane depending on their color and the zenith angle at which they were observed. It manifests as an apparent bulk shift of source position towards the zenith with respect to its actual position, as well as an elongation of the Point Spread Function (PSF) measured over some wavelength range in the resulting images \citep{meyers2015, clarke2024}. DCR occurs because of wavelength-dependent refraction of incident starlight. For example, at airmass $X=1.75$\footnote{Airmass is defined as $X = \sec(Z)$, where $Z$ is the zenith angle}, a source emitting at $\sim5,000\rm\AA$, close to the peak emission of a Sun-like star, is displaced from it's unrefracted position 1.02'' more than a source emitting at $\sim8,500\rm\AA$, the peak emission of an M3.3 dwarf \citep{khata2020, filippenko1982}. Stars at higher air mass experience a stronger color-dependent shift towards zenith. The effect varies across different telescope designs, and is significantly suppressed in telescopes that incorporate atmospheric dispersion correctors (ADC) into their optical path. Yet, use of DCR as a tool for astrophysical discovery has several precedents. Studies such as \citet{kaczmarczik2009} and \citet{richards2018} have demonstrated improvements in redshift estimation of quasars detected in the Sloan Digital Sky Survey (SDSS). \citet{lee2024} showed that DCR could be used to measure astrometric redshifts for simulated Type Ia supernovae in the Vera C. Rubin Observatory Legacy Survey of Space and Time (LSST, \citealt{ivezic2019}) Deep Drilling Fields (DDFs), which when combined with the observed photometric redshifts of these supernovae and that of their host galaxy, could lead to an improvement of the overall redshift estimate. 

Like these examples, observations of stellar flare colors can benefit from an alternative approach when traditional methods are less effective or unavailable. Flares are excellent candidates for atmosphere-aided studies due to the high chromatic contrast to their quiescent counterpart, their increased frequency on low mass stars with convective envelopes, which represent the majority of stars in the galaxy, and their brief duration, which makes detections of the same event in multiple filters difficult. Most importantly, when employed in a large-field synoptic survey such as LSST, atmosphere-aided color-temperature inference unlocks the possibility for an unprecedented population-level survey of stellar flare temperatures. 

Although surveys such as LSST preferentially target low air mass pointings in order to maximize image quality, \citet{clarke2024} notes that in a simulation of the LSST (\texttt{baseline\_v3.0\_10yrs}, \citealt{PSTN-055}), 41\% of all visits in LSST's primary survey mode are performed at $X > 1.25$, which is sufficient to resolve the excess DCR produced by a $10,000$~K flare on an M5 dwarf in the LSST $g$-band. While the cadence of LSST's main survey is slow compared to the average flare duration, such that the event will only be captured in 1-2 datapoints, temporally resolved observations will be available in the LSST DDFs as well as other wide-field surveys Argus Optical Array \citep{law2022}, albeit for a smaller expected number of total flare detections.

In this paper, we present the first unambiguous observation of a flaring source's motion in the zenith direction coincident with its photometric rise, as well as the first DCR-derived characterization of a stellar flare blackbody temperature and filling factor. These data put novel constraints on the physical properties of an M dwarf flare that have applications for flare stars on large scales. We also reflect on the limitations to our forward modeling approach and describe multiple physically-motivated modifications to our initial assumptions that may inform future studies.

This work is reproducible, all material and analysis code used in this work is made available in a GitHub repository.\footnote{\url{https://github.com/RileyWClarke/RAFTS/tree/main/Notebooks/DWF_DCR_flare}}
\newpage
\section{Stellar Flare Temperature Diversity}

Flares are rapid brightening events that occur on stellar photospheres as a result of magnetic reconnection \citep{pettersen1989}. The release of magnetic energy accelerates charged particles into the stellar material, causing localized heating and emission across all wavelengths but most strongly in the blue optical and NUV bands. On stars with cool photospheres such as M dwarfs (\dM\ hereafter), the emitting region contrasts sharply with the redder background, enhancing what is known as the ``visibility'' of the event \citep{gershberg1972}. In photometry, flares appear in two consecutive phases: a fast, highly chromatic initial rise lasting tens of seconds, followed by a more gradual decay phase lasting from minutes to hours \citep{davenport2014,yan2021}. 

Flare temperatures have been an ongoing area of study, using both photometric and spectroscopic approaches. The spectral energy distribution (SED) of stellar flares is dominated by a white light continuum, which is canonically approximated as a constant-temperature ($T_\mathrm{BB} = 9,000 - 10,000$~K) blackbody in studies conducted with broad optical bands (\eg\ \citealt{shibayama2013}, \citealt{osten2015}, \citealt{gunther2020}, etc.). However, numerous studies have revealed significant uncertainty in the effective temperature in these events, demonstrating the continuing need for large scale statistical characterization of flare temperature demographics. \citet{kowalski2013} found, among a sample of simultaneous, high-cadence spectra and photometry of 20 emission-line M dwarfs   (\dMe\ hereafter), continuum emission best fit by blackbody temperatures ranging from $T_\mathrm{BB} \sim 9,000-14,000$~K. \citet{howard2020} used simultaneous Transiting Exoplanet Survey Satellite (TESS, \citealt{ricker2010}) and Evryscope \citep{law2015} data to construct high-cadence time-resolved temperature profiles of 47 ``superflares'' (\ie\ flares with energies $\geq 10^{33}$ erg), 43\% of which reached peak temperatures above the $14,000$~K upper limit found by \citet{kowalski2013}. Additionally, \citet{berger2023} used FUV/NUV flux ratios measured by NASA's Galaxy Evolution Explorer (GALEX, \citealt{morrissey2007}) and found that the assumption of a $9,000$~K blackbody continuum would under-predict FUV flux for 179 out of 183 flares in their sample. Given that flares with FUV temperatures in excess of $40,000$~K have been shown to increase photodissociation rates in exoplanet atmospheres by factors of 10-100 \citep{loyd2018, froning2019}, flares are of significant astrobiological interest and a dominant transient source in large time-domain astronomical surveys that will be crucial tools for characterizing both flare rates and and their effective temperatures.

While missions like \textit{Kepler} \citep{borucki2003} and TESS have expanded flare studies into a new statistical era, these surveys do not provide flare color or temperature constraints. Flare sample sizes using data from these missions range from hundreds to thousands of events per star, \citep{davenport2014, davenport2016, pugh2016, yang2019, gunther2020, pietras2022}, enabling population-level analysis of white light flare observations. However, the total number of flares whose colors or temperatures have been measured directly remains small in comparison to the number of broadband photometric measurements enabled by the continuous, years-long monitoring campaigns of \textit{Kepler} and TESS. The success of wide-field time domain surveys carried out by terrestrial observatories such as the Zwicky Transient Facility (ZTF, \citealt{bellm2019, graham2019}) and the flare catalogs created using the large scale data products they produce (\eg\ \citealt{crossland2024,
voloshina2024}) represent an important complement to space-based flare observations. The next generation of astrophysical time domain surveys, \eg\ LSST \citep{ivezic2019}, with their promise to deliver unprecedented volumes of data, likewise herald a new era in time domain studies of stellar flares. However, the scientific goals of these surveys require observing strategies that are less favorable to traditional modalities of flare studies, meaning that innovative methods must be developed and deployed in order to leverage the massive increase in data that will soon be available.



\section{DECam and the Deeper, Wider, Faster Programme}

The Dark Energy Camera (DECam, \citealt{flaugher2015}), mounted on the 4-meter Victor M. Blanco Telescope on Cerro Tololo, Chile, is one of the optical components of the \textit{Deeper, Wider, Faster} (DWF) Programme \citep{andreoni2017}, which coordinates over 80 observatories worldwide to collect simultaneous, multimessenger, multiband observations of transients on milliseconds-to-hours timescales. Running DECam at a $\sim$50 seconds cadence\footnote{20s exposures separated by $\sim$30s read out \citep{andreoni2020}}, DWF data probes minute timescales, accessing a region of parameter space difficult to reach with other terrestrial transient surveys. By exploring the rapid transient sky with a limiting magnitude of $\sim$23 in $g$, DWF is able to detect and characterize a broad range of variable phenomena, including flares (\eg\ \citealt{webb2020}, \citealt{andreoni2020}).

In a volume-limited study of stellar flares using DWF DECam data, \citet{webb2021} used unsupervised machine learning to detect 96 flare events on 80 stars across 12 DWF fields. One of the most energetics among these events, DWF030225.574-545707.456, 
occurred on the M7 dwarf Gaia DR2 4728703055241994752 on 12/18/2015 06:10:38 UTC. 
This event was not only bright 
($\Delta m_g$ = -6.12; for comparison, a flare on an M3 dwarf with an assumed $9,000$~K blackbody SED would produce an observed $\Delta m_g$ = -0.3; see \citealt{davenport2012}) but also 
observed at a high range of air masses ($1.45 \lesssim X \lesssim 1.75$). While prioritizing observations at low airmass maximizes image quality and depth,
coordinating simultaneous observations between telescopes located in Chile and in Australia require some fields be observed when they are at larger zenith angles and thus greater airmass. This benefits any study that wishes to leverage DCR as a tool, as the positional offset is proportional to air mass. Furthermore, the flare was observed using DECam's $g$ filter, which is the optimal passband for measuring the difference in DCR before and during the flare ($\Delta$DCR) to infer the effective flare temperature \citep{clarke2024}. The high brightness, visibility, air mass, and favorable passband at which this particular event was observed, combined with the subsecond resolution of the DECam images (0.26 arcsec/pixel pixel scale), makes it an ideal case study for atmosphere-aided flare temperature inference.







\section{Measuring the observable properties of DWF030225.574-545707.456}\label{sec:astrometry}

The data used in the following analysis was collected in a single DECam run on 12/18/2015 observing a 3 square degree field centered on RA=03:00:00 and Dec=--55:25:00 (J2000). The data was taken using DECam's $g$ filter, with an expected limiting magnitude of $g\lesssim$ 23. A total of 74 exposures were taken over a period spanning 0.98 hours.  

\begin{figure}
     \centering
     \includegraphics[width=0.95\linewidth]{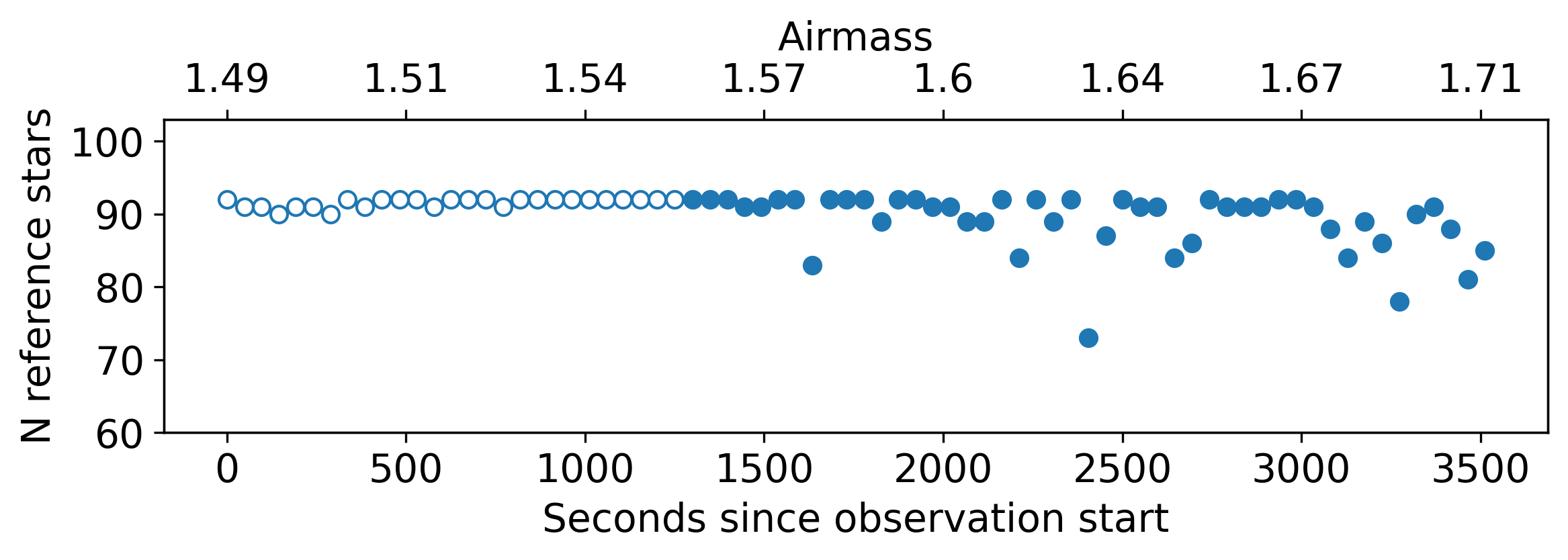}
     \caption{Number of reference stars detected in each frame. Empty circles are pre-flare, filled circle are after the start of the flare (see \autoref{sec:astrometry}). Between 72 and 94 stars are detected and used as reference in each frame.}
     \label{fig:refstars}
 \end{figure}

We performed PSF photometry using \texttt{Photutils} \citep{photutils} on the image files corresponding to the CCD in which the flare was detected (CCD S18). Source detection was performed using \texttt{DAOStarFinder}, part of the \texttt{photutils.detection} subpackage that implements the \texttt{DAOFIND} algorithm \citep{stetson1987}. Using the 7th image in the full sequence (timestamped 2015-12-18T06:33:52.769) we find a list of 92 stars, including the flare star. The 91 reference stars were chosen such that they were reasonably well distributed across the entire CCD as well as to avoid saturated sources.\footnote{The DAOPhot parameters used were \texttt{DAOStarFinder} parameters used: \texttt{roundlo} = -0.5, \texttt{roundhi} = 0.5, \texttt{peakmax} = 10,000, \texttt{exclude\_border}=\texttt{True}, and FWHM set to the image header seeing value of $FWHM\sim~4.9$~ pixels} The field is centered at a galactic latitude of $b\sim-55\deg$, thus is not particularly crowded. However, due to the relatively poor seeing and high airmass that increases over the observing run, the quality of the images varies throughout the time series. 
With this starting point, \texttt{DAOStarFinder} is applied to each frame with the same parameters except \texttt{FWHM}, which is increased to up to $3\times$ the image seeing as estimated in the fits header, until the flare star Gaia DR2 4728703055241994752 is detected. Photometry is then performed on all detected sources with the \texttt{PSFPhotometry} class from the \texttt{Photutils} \texttt{Python} package \citep{larry_bradley_2024_13989456}. We tested a variety of PSF models, including a Moffat profile and an empirical PSF generated from bright stars in each image, and found that the most stable results were achieved with a 2D Gaussian model as implemented in the \texttt{GaussianPRF} class with the 2D FWHM fit to the data independently in each dimension.\footnote{\new{For every image in the sequence, we run \texttt{PSFPhotometry}, initializing the search with the star finder generated  \texttt{DAOStarFinder}  and use an instance of the \texttt{GaussianPRF} where \texttt{x\_fwhm.fixed} \texttt{y\_fwhm.fixed} are set to \texttt{False} as PSF model to enable independent fit of the PSF shape on the $x$ and $y$ axis, corresponding to a (potentially) elliptical PSF.}} Of the 91 reference stars, 34 \new{are detected in all frames and at least 72 stars are detected in each frame (\autoref{fig:refstars}). These} reference stars allow us to construct trend time series (constructed as the weighted average of all star measurements, weighted by the standard deviation of a star across the time series). We thus obtain time series of photometry (flux) and astrometry (RA and Dec centroids) for all 92 stars, with associated uncertainties, which we use for the subsequent analysis.\footnote{\new{We tested the robustness of our result to different approaches in constructing trends using the mean position, the median position, and the weighted mean position of reference stars, and using either all detected stars or only the stars detected in all frames; we found that our derived astrometry was insensitive to these choices, with only negligible differences.}}
We define the start of the flare when Gaia DR2 4728703055241994752's PSF magnitude rises $g>\new{3}\sigma$ above the median flux of the first $1,000$ seconds of observation, where $\sigma$ is the standard deviation of the magnitude calculated over the same period, or MJD=57374.256832.\footnote{Arguably, the brightening began a few time stamps before but we opt for a conservative estimate.}

\begin{figure}[!ht]
    \centering
    \includegraphics[width=0.475\textwidth]{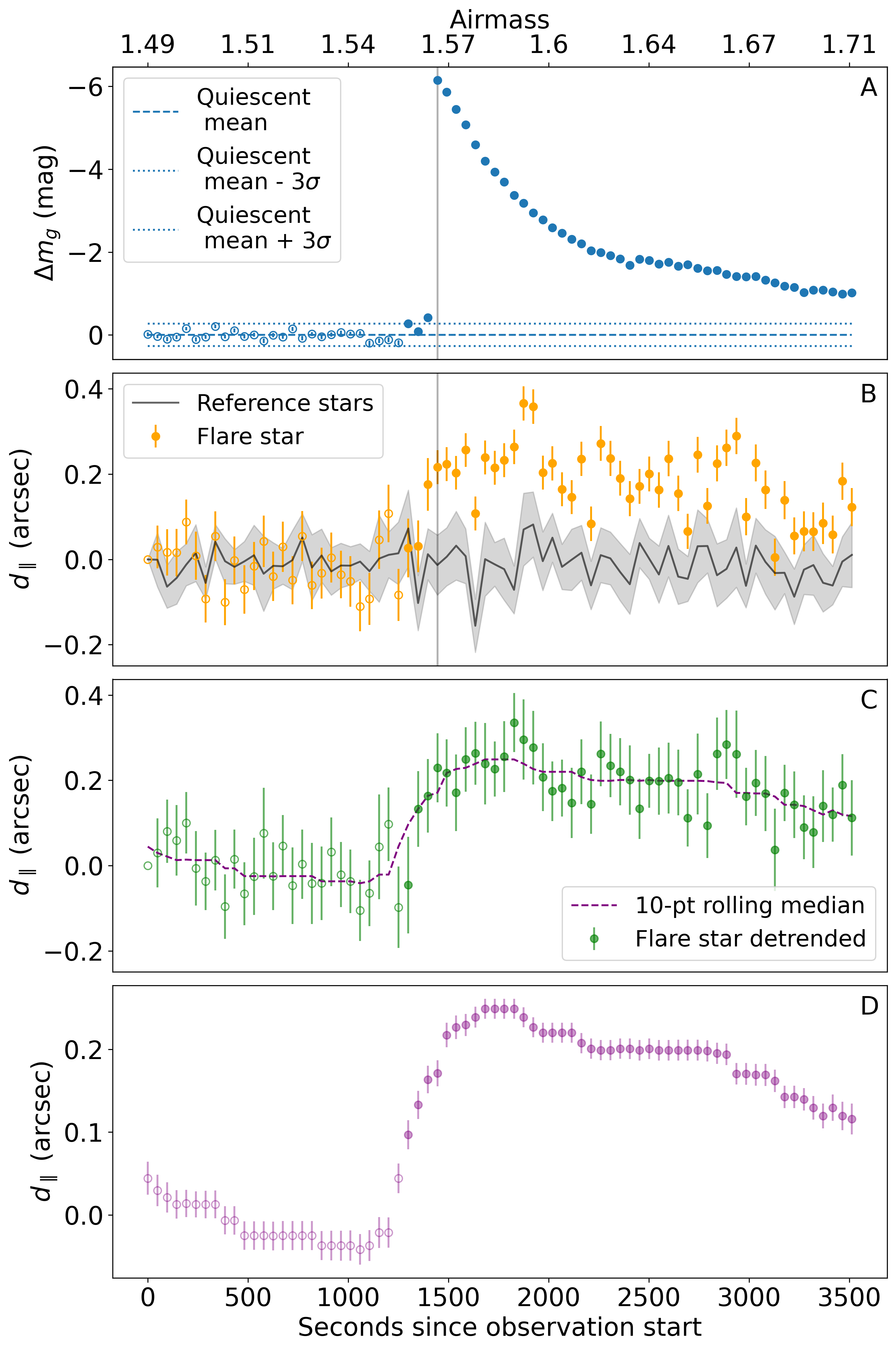}
    \caption{\textit{Panel A}: Light curve of the DWF030225.574-545707.45 flare expressed as $g$ magnitudes in excess of the quiescent stellar brightness. \new{Hollow points indicate pre-flare epochs and solid points indicate flaring epochs. We define the start of the flare as the first measurement to exceed the quiescent magnitude by $3\sigma$, where the quiescent magnitude is taken to be the mean magnitude of the first 1000 seconds of observation, and $\sigma$ is the standard deviation of all points within the first 1000 seconds.} As noted in \citet{webb2021}, the lightcurve does not return to pre-flare brightness before the end of the observing period. \textit{Panel B}: Raw $d_\parallel$ measured on the images, relative to source position at $t_0$. The weighted average $d_\parallel$ of 91 reference stars, weighted by the stars' astrometric standard deviation across the entire time series, is shown as a grey line and the standard deviation of their relative position itself is shown as a filled region. \textit{Panel C}: detrended $d_\parallel$ of the flare star generated by subtracting the weighted average $d_{\parallel}$ of the reference stars shown in Panel B. A 9-point rolling median is shown by the dashed black line. \textit{Panel D}: Rolling median over an 9-point window of the detrended data shown in Panel C, and the time series used for subsequent analysis in this work.}
    \label{fig:dpar}
\end{figure}

At $\sim$1,500 seconds into the observing period, the \dM\ Gaia DR2 4728703055241994752 reaches a peak brightness of $\Delta m_g$ = --6.13 mag, as measured from the median magnitude of the first $1,000$ seconds of observations. The star does not return to its quiescent magnitude before the observing period ends, thus the full duration of the flare event is unknown, but the lower bound on the duration is 2115.37 seconds, over which period the $g$ band energy released was measured to be at minimum $2.38\times10^{35}$ ergs \citep{webb2021}. The lightcurve of the event is shown in \autoref{fig:dpar} (top panel). 

Source centroids and centroid errors were converted from pixel coordinates to world coordinates using the Astropy WCS library \citep{astropy}.
We measured the change in world coordinates ($\Delta\alpha$, $\Delta\delta$) for all 91 sources across all 74 exposures, relative to their positions at the start of the observing sequence ($t_0$ hereafter) in order to compare the flare star's apparent positions to that of the other stars in the images. We calculate the parallactic angle of each source at each epoch, defined as: 
\begin{equation}\label{equation:pa}
P=\tan^{-1}\left(\frac{\sin (h)}{\cos (\delta) \tan (\phi)-\sin (\delta) \cos (h)}\right),
\end{equation}
where $\delta$ is the object's declination, $h$ is the object's hour angle, and $\phi$ is the geographic latitude of the observer's location on Earth. $\Delta\alpha$, $\Delta\delta$, and $P$ are then used as inputs to calculate $d_{\parallel}$ as:
\begin{equation}\label{equation:dpar}
d_{\parallel}=\sqrt{\Delta \alpha^2+\Delta \delta^2} \cos \left(\frac{\pi}{2}-P-\arctan \frac{\Delta \delta}{\Delta \alpha}\right),
\end{equation}
which is the projection of the source's apparent positional deviation in the direction of the zenith and serves as a quantification of \ddcr\ \citep{clarke2024}. The weighted average $d_{\parallel}$ of the reference stars (weighted by the standard deviation of each star time series) is then subtracted from that of Gaia DR2 4728703055241994752 (\autoref{fig:dpar} Panel C). The detrended time series is then smoothed via a 9-point rolling median (9 chosen empirically as the smallest window that effectively removes short time scale noise), as shown in \autoref{fig:dpar} Panel D. \new{While the curve in Panel D contains one epoch that appears to increase just prior to the start of the flare, indicated by the solid points, this is an artifact of the rolling window applied to the data in Panel C and should not be interpreted as pre-flare $\Delta$DCR.}

We note the seeing FWHM on this particular night was poorer than typical ($FWHM\sim1.33$~arcsec), and CTIO regularly achieves $<$ 1 arcsec FWHM seeing at similar air masses, suggesting that observations under more favorable conditions would provide superior PSF centroid estimation, which would enable $\Delta$DCR shift estimation for weaker flares.

\subsection{Statistical Significance of the Change in $d_\parallel$}\label{sec:statanalysis}

On the detrended $d_\parallel$ time series (\autoref{fig:dpar} Panel C), we run a point of change (POC) analysis to assess the statistical significance of the $\Delta$DCR signal (for a review of POC analysis methods, see \citealt{basseville1993detection}). With a standard off-line POC detection approach, we search for a split in the time series, a POC, that maximizes jointly the difference between means and the $\chi^2$ before and after the POC. We only search points between MJD=57374.247350 and MJD=57374.278546, avoiding the first and last five points in the time series to avoid excessively short segments. We find the most likely POC in $d_\parallel$ at MJD = 57374.257382, corresponding to the peak of the flare. This POC is visualized as a vertical line in  \autoref{fig:dpar} Panels A and B. This method searches for a single POC, but we note that running the standard Python package \texttt{ruptures} we repeated the POC detection with the algorithm suggested in \citealt{celisse2018new} with a Gaussian kernel $k = e^{-\gamma(\mu_{t-1:t})}$ we also find a single POC for any value of the hyperparameter $\gamma<7$.

We then assess the statistical significance of the POC we detected. A Mann-Whitney U test \citep[a non-parametric test][]{mcknight2010mann} indicated that the before- and after-POC $d_\parallel$ segments are statistically inconsistent at $>5\sigma$ ($U= 34$). 
Further analysis requires assuming a model for the before- and after-POC $d_\parallel$ time series. For simplicity, we assume piecewise-stationary Gaussian $\mathcal{N}(\mu_i,\sigma_i)$ generative models, one with two stationary segments with mean the mean of the segment  ($\mu_0=<d_\parallel[0...POC)>$ and $\mu_1=<d_\parallel[POC...N)>$), and one with a single segment with mean $\mu=<d_\parallel>$, the mean of the entire time series, and $\sigma$'s the average uncertainties of each segment.\footnote{{We also tested using the segments' standard deviation, which leads to the same statistical conclusions.}} We find a Bayes odds ratio $Br\sim10^{21}$, indicating extremely strong statistical preference for the piecewise model (well in excess of the canonical 150 value for decisive significance \citep{jeffreys1998theory}), and a likelihood ratio test indicates the piece-wise model is preferred at a $>5\sigma$. Finally, we note that the generative models we adopted are not realistic, as the $d_\parallel$ should evolve during the flare and eventually return to baseline, but we prefer a simple model at this stage of the analysis, and a more accurate description of the $d_\parallel$ flare evolution can only result in models that would be even more strongly favored by statistical tests.

We note that the light curve (Panel A in \autoref{fig:dpar}) and the time series of $d_\parallel$ (Panel D) have distinct evolutions. This is not unexpected. For example, the delayed onset of Ca II K line emission has been well documented in the literature  \citep{hawley1991, houdebine1991, houdebine2003, garcia2002, kowalski2013}. Line emission produces a proportionally larger change in effective wavelength compared to brightness, so the delayed line emission could produce a positional shift even after the flare brightness has decayed. This is further discussed in \autoref{sec:lines} and \autoref{sec:conclusions}.

\section{Modeling a flare's temperature evolution } \label{sec:analysis}

In this section, we describe all the elements of the models that, combined, will allow us to infer the temperature of a flare from single-band photometry: the quiescent source characteristics, the relationship between flare observables (magnitude and color) with the flare temperature and filling factor. 
\subsection{Quiescent source}\label{sec:quiescent}

The observational properties of DWF030225.574-545707.456 are reported in \citealt{webb2021}.\footnote{Apparent magnitude in quiescence $g=20.94$, and the survey expected limiting magnitude in $g$ band is $m(AB) \sim 23$, for an average seeing of 1.0 arcseconds and airmass of $X=1.5$, see \citealt{webb2021} for more details.} To model the flare, we need to first constrain the quiescent SED. Based on its DR3 Gaia magnitude, color, and parallax ($M_G = 18.1124 \pm 0.0016$ mag, $G_{BP} - G_{RP} = 3.9014 \pm 0.0685$ mag, p = 10.18 mas, respectively) we reclassify the flare star Gaia DR2 4728703055241994752 as an M7 dwarf (\dM7) \citep{gaia2016, gaia2021} and use a template spectrum of an active \dM7 as the SED of the quiescent source (the source was originally reported as an \dM5 in \citealt{webb2021} using the Gaia DR2 color-magnitude values). 

\subsection{Simultaneous Forward Modeling of Temperature and Filling Factor}\label{sec:forwardmodel}

In \citet{clarke2024}, we showed how modeling a flare as a blackbody at a specific temperature ($T_\mathrm{BB}$) and with assumptions on filling factor ($X_\mathrm{BB}$) we could predict the apparent zenith-ward positional shift of a flaring star. 
The filling factor is the 2-dimensional projected flaring area as a fraction of the total visible stellar hemisphere and it is treated as a linear scaling of the flaring spectrum. The flaring spectrum itself is calibrated such that a $10,000$~K 
blackbody with a fiducial filling factor of 0.25\% has the same 
total radiance as the quiescent star over the SDSS optical range 
($3,980 - 
9,200~\rm\AA$). 
The calibration values were selected to be representative of temperature-filling factor combinations observed in studies such as \citet{kowalski2013, kowalski2016}. Together, $T_\mathrm{BB}$ and $X_\mathrm{BB}$ determine the contribution of the flare to the star's SED.

To quantify the color change of a source, and thus its temperature, we model the effective emission wavelength, $\lambda_\mathrm{eff}$ as a function of $T_\mathrm{BB}$ and $X_\mathrm{BB}$. In our model, $\lambda_\mathrm{eff}$ is defined by:
\begin{equation}
    \lambda_\mathrm{eff} = \mathrm{exp}\left[\frac{\int_0^\infty f_\lambda(\lambda, T_\mathrm{BB}, X_\mathrm{BB}) S_g(\lambda)\ln(\lambda)d\lambda}{\int_0^\infty f_\lambda(\lambda, T_\mathrm{BB}, X_\mathrm{BB}) S_g(\lambda)d\lambda}\right],
    \label{eq:weff}
\end{equation}
where $S_g$ is the transmission function of the DECam $g$ filter 
and $f_\lambda$ is the spectral flux of the source SED in units of $Wm^{-2}\mu^{-1}$. The uncertainties on the measured source centroids ($\alpha$, $\delta$) are propagated to the effective wavelength ($\lambda_\mathrm{eff}$) at each epoch by sampling, with replacement, 1,000 values for each of RA and Dec from a normal distribution centered on the measured coordinates and with standard deviation given by their uncertainty (implicitly assuming that the errors are Gaussian and independent, which provides a conservative estimates for the uncertainty), and recalculating $\lambda_\mathrm{eff}$ 1,000 times per epoch. At each epoch, the uncertainty in $\lambda_\mathrm{eff}$ is then given by the standard deviation of the distribution of 1,000 values thus obtained. 

The magnitude change $\Delta m_g$ relates to $T_\mathrm{BB}$ and $X_\mathrm{BB}$ as follows:
\begin{equation}
    f_{\lambda, g} = \frac{2\pi R_*^2}{d^2} ~\frac{\int_0^\infty f_\lambda(\lambda, T_\mathrm{BB}, X_\mathrm{BB}) S_g(\lambda)d\lambda}{\int_0^\infty S_g(\lambda)d\lambda} ,
    \label{eq:flux}
\end{equation}
where $f_{\lambda, g}$ is the $g$ band spectral flux density, $R_*$ is the stellar radius and $d$ is the distance to the star. The source magnitude is then  

\begin{equation}
    m_g = 22.5 - 2.5 * \log_{10}(f_{\lambda, g})
\end{equation}
and $\Delta m_g$ is the difference between $m_g$ calculated using the flaring spectrum versus the quiescent spectrum. In both \autoref{eq:weff} and \autoref{eq:flux}, the spectrum $f_\lambda$ is convolved with the DES $g$ band transmission function $S_g(\lambda)$, which includes atmospheric extinction for an $X = 1.2$ atmosphere.


\subsection{Instrumental effects} \label{sec:chrdist}

In addition to DCR, any axisymmetric optical system such as DECam is expected to have color-dependent radial distortion described by odd-ordered polynomials in distance from the field center \citep{bernstein2017}.\footnote{Here, ``field center'' refers to the center of DECam's 3 sq. deg. field of view, and not the CCD center.} This manifests as an additional chromatic term contributing to the observed positional offset $d_\parallel$. We correct for this effect using data produced by a simulation of the DECam optical system using \texttt{batoid} \citep{meyers2019}, an open-source, optical ray-tracing software package. 
These data 
describe the relative focal plane position of a source as a function of wavelength over a range of $4,000-5,500~\rm\AA$, and of field angle from 0.0 to 1.1 deg.  As the field was tracked by DECam over the entire observing period, the field angle of the flare star differs from its mean value of 0.5793 deg by no more than 0.45 arcseconds. 
Therefore, we slice this data at a fixed field angle, and use the relative focal plane position as a function of wavelength only to estimate the magnitude of the distortion. We fit a quadratic polynomial to the data as a function of wavelength to derive the radial contributions to the shift of our star. 


Because this distortion shifts the source radially rather than toward zenith, we calculate the contribution of this effect to the measurement of $d_\parallel$ by taking the projection of the radial distortion onto a unit vector pointing to the zenith. This is done by taking the great circle distances from the field center to the source and from the source to zenith, respectively, and using the spherical law of cosines to find the angle between the two great circles.

\begin{figure*}[!ht]
    \centering
    \includegraphics[width=0.9\textwidth]{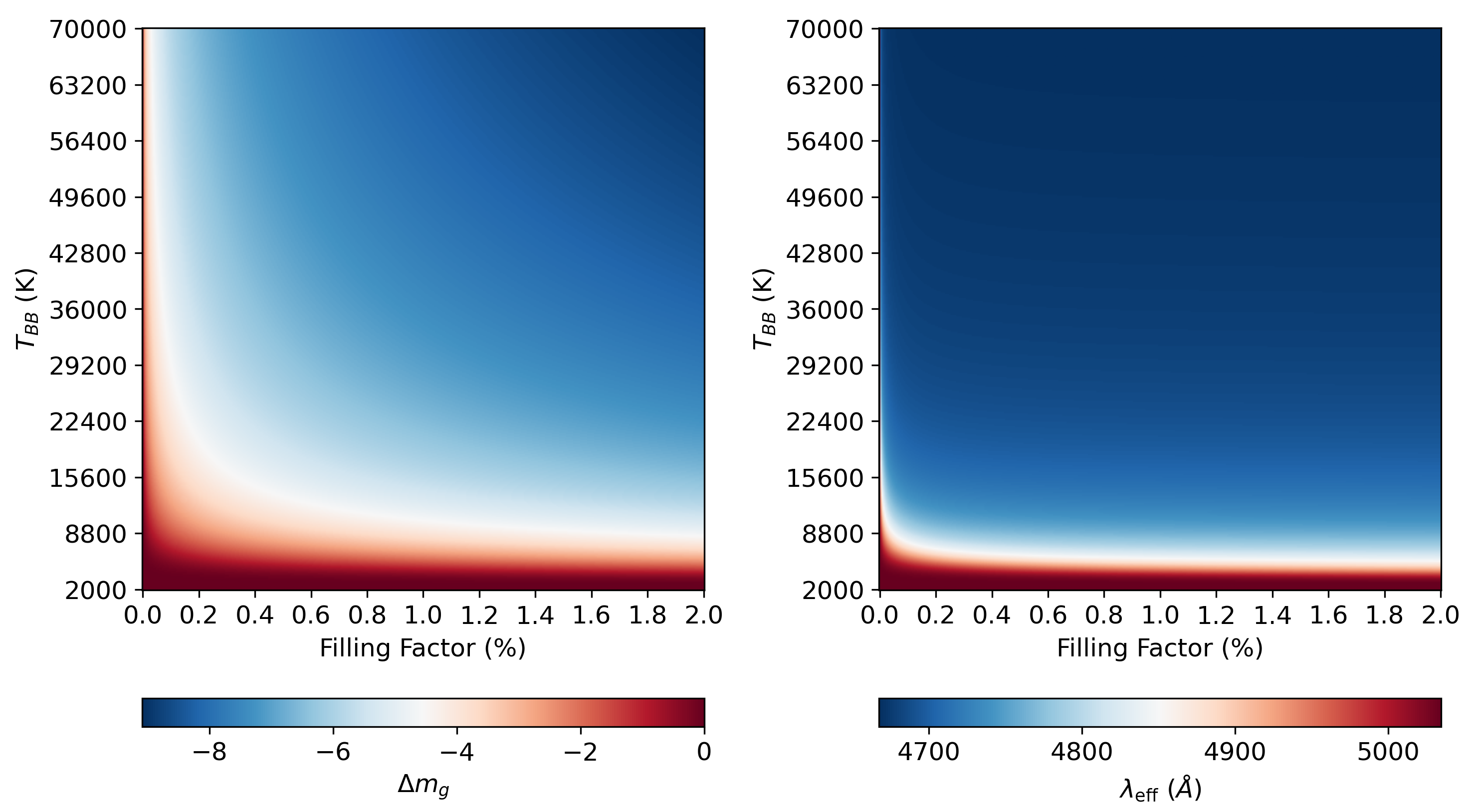}
    \caption{Left: Change in $g$ magnitude relative to the quiescent stellar magnitude as a function of flare temperature and filling factor. Right: Effective wavelength of the model flare spectrum in the $g$ band as a function of flare temperature and filling factor. Each grid consists of 400x400 cells, giving a resolution of 170 K in temperature and 0.005\% in filling factor.
    }
    \label{fig:grids}
\end{figure*}

The great circle distance is defined as 
\begin{equation}
    \sigma=\arccos \left(\sin \delta_1 \sin \delta_2+\cos \delta_1 \cos \delta_2 \cos \Delta \alpha\right)
    \label{equation:gcd}
\end{equation}
where $\alpha_1$ and $\delta_1$ are the RA and Dec of the first position, $\alpha_2$ and $\delta_2$ are that of the second position, and $\Delta \alpha = |\alpha_1 - \alpha_2|$.

Let $\sigma_{fz}$, $\sigma_{sf}$, and $\sigma_{zs}$ be the great circle distances between field center--zenith, flare star--field center, and zenith--flare star, respectively. By the spherical law of cosines, the angle $\Theta$ between two great circles, one passing through the flare star and the field center and one passing through the flare star and the field center is

\begin{equation}
    \Theta = \cos^{-1}\left(\frac{\cos(\sigma_{fz}) - \cos(\sigma_{sf})\cos(\sigma_{zs})}{\sin(\sigma_{sf})\sin(\sigma_{zs})}\right)
\end{equation}

The amount of radial distortion in the direction of the zenith is then quantified by $R_\mathrm{dist}(\lambda_\mathrm{eff})\cos(\Theta)$, where the magnitude of the distortion $R_\mathrm{dist}$ is described as:

\begin{equation}
    R_\mathrm{dist} = -5.26*10^{-8}(\lambda_\mathrm{eff})^2 + 5.86*10^4(\lambda_\mathrm{eff}) - 1.58
\label{Rdist}
\end{equation}
which is a second degree polynomial interpolation to measured $R_\mathrm{dist}-\lambda$ pairs. At a field angle of $0.5793^{\circ}$, for an M7 flare with representative temperature $T_{BB} = 10,000~K$ ($\lambda_\mathrm{eff}=4757.7$) in $g$ band, see \citealt{clarke2024}) the effect is $R_\mathrm{dist}\sim0.014$ arcsec, about 3\% of the observed positional offset, and for a very hot flare of $T_{BB} = 40,000~K$ ($\lambda_\mathrm{eff}=4675.9$), the maximum temperature we will measure for this flare, see \autoref{sec:weffs}), the instrumental offset shrinks to $R_\mathrm{dist}\sim$0.007 arcsec.

\subsection{Temperature and Filling Factor Derivation} \label{sec:weffs}

Now that we have all elements of our model, each $d_\parallel$ measured on the images is converted to $\lambda_\mathrm{eff}$ via the following procedure:

\begin{enumerate}
    \item The \ddcr\ measured on the image ($d_{\parallel}$) is defined as
    \begin{equation}
    \begin{split}
        d_\parallel = & R_\mathrm{flare}(\lambda_\mathrm{eff}) - \\& R_\mathrm{dist}(\lambda_\mathrm{eff})\cos(\Theta) - \\
        & R_\mathrm{quiescent},
        \label{eq:refraction}
    \end{split}
\end{equation}
where $R_\mathrm{flare}$ is the angular distance of the source from its unrefracted position during the flare in arcseconds,  $R_\mathrm{dist}\cos(\Theta)$ is the color-dependent radial distortion correction described in \autoref{sec:chrdist}, $R_\mathrm{quiescent}$ is the angular distance of the source from its unrefracted position during quiescence in arcseconds, and $\lambda_\mathrm{eff}$ is the effective wavelength of the source.
\item DCR at quiescence ($R_\mathrm{quiescent}$) is estimated using an quiescent \dM 7 template spectrum built using SDSS observations \citep{bochanski2007}. Notice that while this term is included explicitly in the equation, the absolute DCR cancels out between the first and third term of \autoref{eq:refraction} as the quiescent source SED is included in both terms.
\item \autoref{eq:refraction} is solved for $\lambda_\mathrm{eff}$ in the terms $R_\mathrm{flare}$ and $R_\mathrm{dist}$ via numerical optimization. The terms $R_\mathrm{flare}$ and $R_\mathrm{quiescent}$ are described by 
\begin{equation}
    R = R_0(\lambda_\mathrm{eff})\tan(Z)
\end{equation}
\begin{equation}
    R_0(\lambda_\mathrm{eff})=\frac{n_\lambda^2-1}{2 n_\lambda^2}
\end{equation}
\begin{equation}
\begin{split}
[n_\lambda - 1]~10^6 = 64.328\: + & \frac{29498.1}{146-(1/\lambda_\mathrm{eff})^2} \\
                              + & \frac{255.4}{41 - (1/\lambda_\mathrm{eff})^2};
\end{split}
\end{equation}
\\
\noindent where $n_\lambda$ is the index of refraction, $Z$ is the zenith angle and $X$ the air mass. 
\end{enumerate}

Using these equations, $\lambda_\mathrm{eff}$ and $\Delta m_g$ can be calculated over a grid of $T_\mathrm{BB}$ and $X_\mathrm{BB}$, shown in \autoref{fig:grids}. These model grids are used as a look-up table in a simultaneous ordinary least squares regression to derive $T_\mathrm{BB}$ and $X_\mathrm{BB}$ by minimizing the objective function $F_\mathrm{obj}$ defined as:


\footnotesize
\begin{equation}
F_\mathrm{obj} = \sqrt{
     \left(\frac{\lambda^\prime_\mathrm{eff} - \lambda^\prime_\mathrm{eff, model}}  {\sigma_{\lambda^\prime}}\right)^2 +
     \left(\frac{\Delta m^\prime_\mathrm{g} - \Delta m^\prime_\mathrm{g, model}}{\sigma_{m^\prime_\mathrm{g}}}\right)^2
}
\label{eq:objective}
\end{equation}
\normalsize

\noindent where the primed quantities indicated that $\lambda_\mathrm{eff}$ and $\Delta m_g$ are scaled to the range [0-1] before being used in the loss function and $\sigma_{\lambda^\prime}$ and $\sigma_{m^\prime_g}$ are the uncertainty in (scaled) effective wavelength and magnitude, respectively (see \autoref{sec:forwardmodel}). 

\begin{figure*}[!ht]
    \centering
    \includegraphics[width=0.85\textwidth]{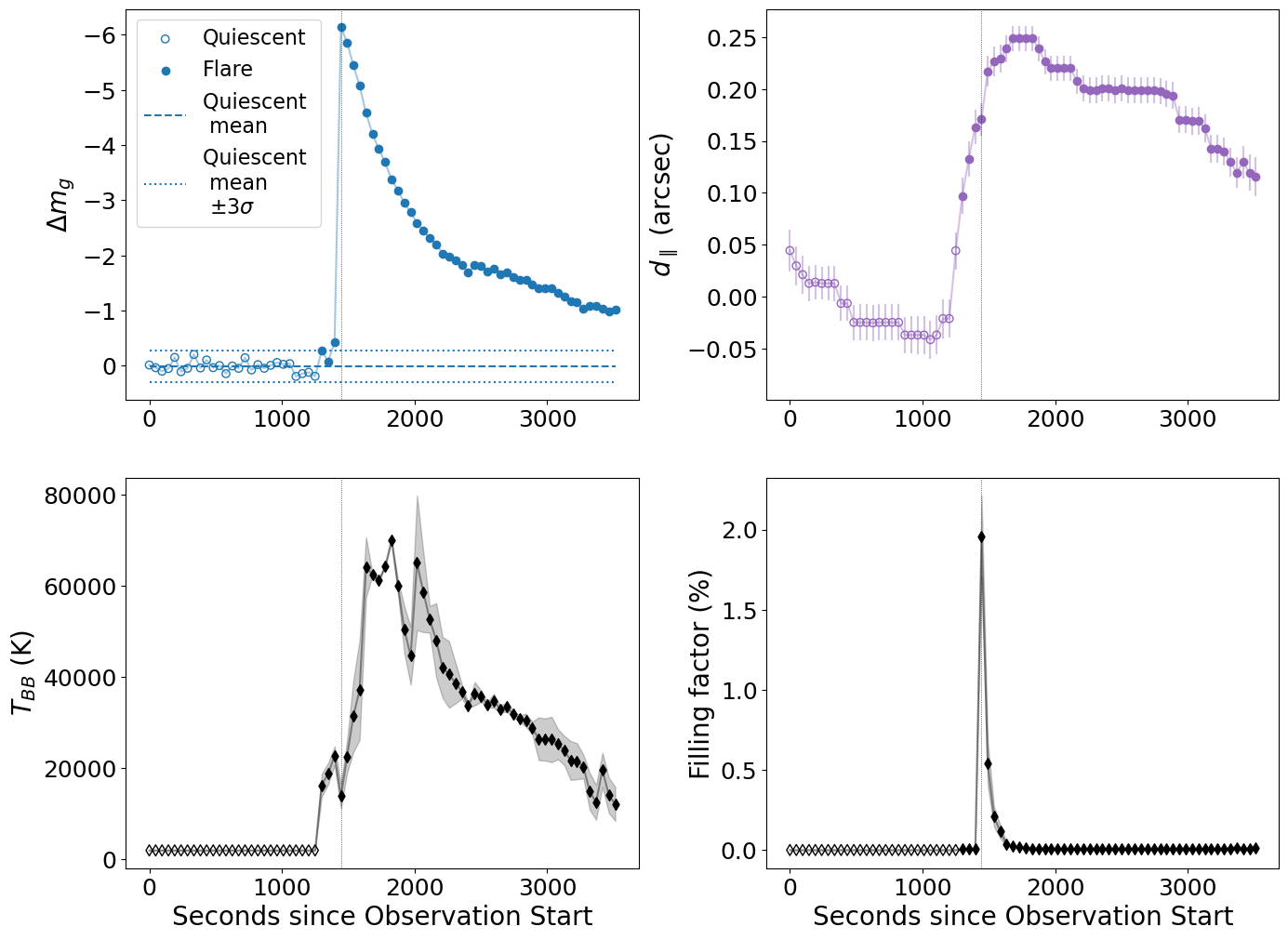}
    \caption{Top left: Flare lightcurve as shown in the top panel of \autoref{fig:dpar}. Top right: $d_\parallel$ as shown in the bottom panel of \autoref{fig:dpar}. Bottom:  the evolution of temperature ($T_\mathrm{BB}$, left) and filling factor (right) from our most simplistic model described in \autoref{sec:weffs}: a blackbody flare on top of a \dM\ SED and no constraints on the filling factor evolution. The peak of the flare is marked in each panel with a vertical line. The uncertainties are derived by bootstrapping over the observable uncertainties 1,000 times.}
    \label{fig:result1}
\end{figure*}

First, $T_\mathrm{BB}$ and $X_\mathrm{BB}$ are derived 
at each step $t$ in the time series for each pair of observables $\Delta m_g(t)$ and $\lambda_\mathrm{eff}(t)$. This result is shown in the second row of \autoref{fig:result1}. For the pre-flare epochs, indicated as hollow points in \autoref{fig:result1}, we fix filling factor to zero, since before the flare occurs the physical extent of the blackbody emission is effectively zero. However, the model as explained thus far contains two naive assumptions: 1) The flare-only SED is well approximated solely by a blackbody and 2) a lack of constraint on how rapidly $X_\mathrm{BB}$ can change. Therefore, we consider two physically-motivated modifications to this procedure: 1) the inclusion of emission features in the flare-only model SED and 2) constraints on the temporal evolution of $X_\mathrm{BB}$.

\subsection{Effects of Emission Features in Flare SEDs} \label{sec:lines}


It is well known that observed flare SEDs regularly diverge from a pure blackbody representation and several spectral features are commonly observed in the wavelength ranges relevant to this study (\eg~\citealt{hawley1991, houdebine1991, houdebine2003, kowalski2013, kowalski2019b, namekata2020}). Therefore, we illustrate in this section how emission features shift the effective wavelength of the flaring SED toward the blue end of the $g$ filter, resulting in a significant increase in $\lambda_\mathrm{eff}$ versus a relatively smaller increase in total flux, implying that modeling the flare solely as a blackbody can overestimate the temperature of the flare.

\begin{figure*}[!ht]
    \centering
    \includegraphics[width=0.95\textwidth]{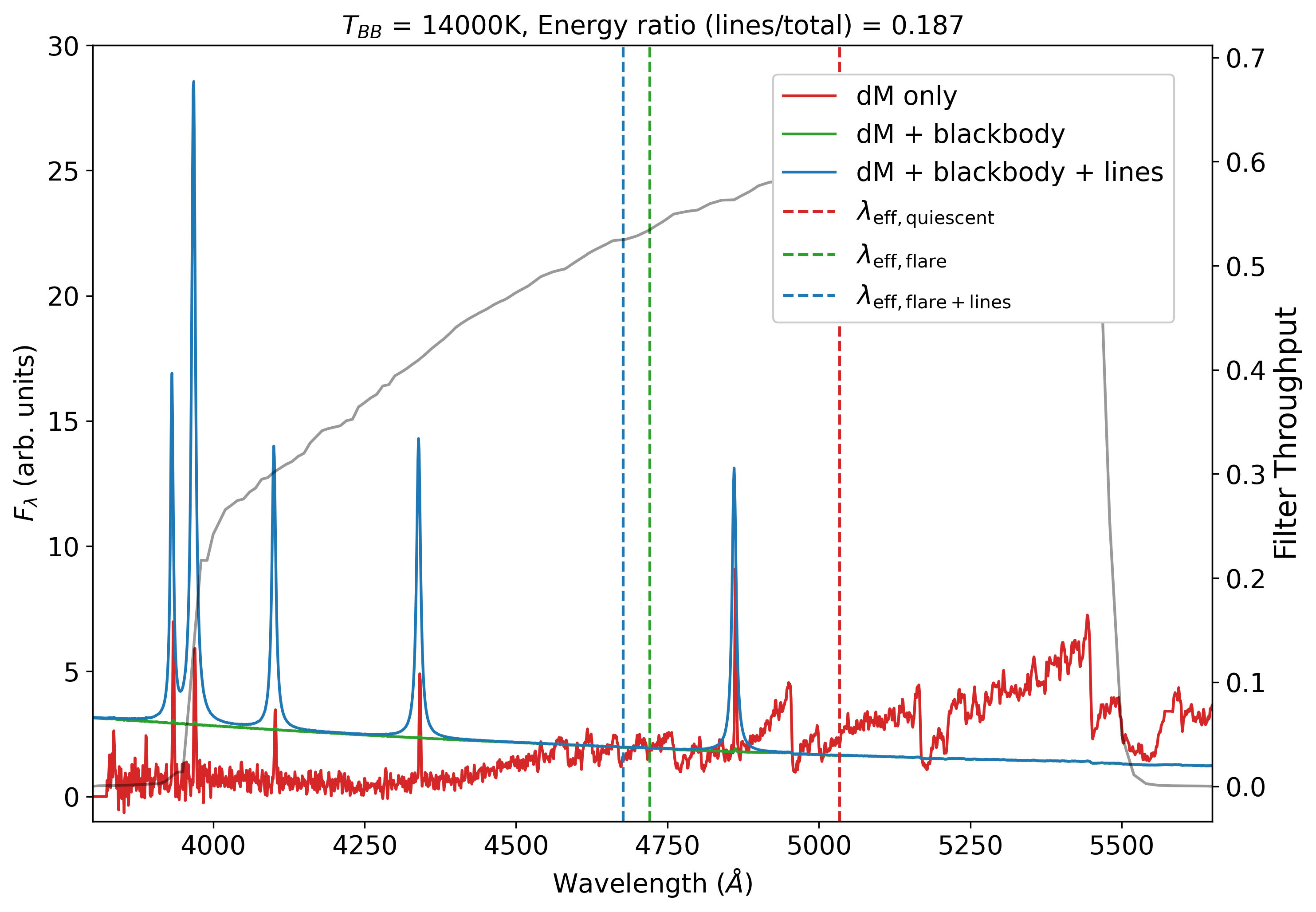}
    \caption{The effect of emission lines on the effective wavelength $\lambda_\mathrm{eff}$ in a stellar flare. In red, the quiescent spectrum of an active \dM7 star (see \autoref{sec:quiescent}). The quiescent spectrum has been magnified by a factor of 60 for visibility. In green, the blackbody-only flaring spectrum with $T_\mathrm{BB} = 14,000$~K and $X_\mathrm{BB} = 0.25\%$. 
    In blue, the flaring spectrum with $T_\mathrm{BB} = 14,000$~K and $X_\mathrm{BB} = 0.25\%$, as well as enhanced Ca H \& K, H$\beta$, H$\gamma$, and H$\delta$ lines. The transmission function of the DES $g$ filter is shown in grey. The vertical dashed lines correspond to the effective wavelength in $g$ for each model: $\lambda_\mathrm{eff, quiescent}=5,035 \rm\AA$, $\lambda_\mathrm{eff, flare}=4,721 \rm\AA$, $\lambda_\mathrm{eff, flare+lines}=4,677 \rm\AA$. While this is a simplistic model for the inclusion of emission lines, it illustrates the effect lines have 
    on $\lambda_\mathrm{eff}$: for the same blackbody temperature, the effective wavelength decreases. Conversely, a smaller temperature is needed to explain the DCR if emission features are enhanced during the flare.}
    \label{fig:lineplot}
\end{figure*}

\autoref{fig:lineplot} illustrates the shift in effective wavelength caused by adding emission-like features to a $14,000$~K blackbody. The lines shown in \autoref{fig:lineplot} are Lorentzian functions fit to Ca II H, Ca II K, H$\beta$, H$\gamma$, and H$\delta$ lines from a flare SED observed by \citet{kowalski2013} using the Dual-Imaging Spectrograph on the 3.5m ARC telescope at Apache Point Observatory. The Lorentzians are then uniformly scaled such that the ratio of energy emitted by the lines in DES $g$ (398-548nm) to that of the total SED is 0.187, which is similar to that observed in some high energy flares such as the Great Flare of AD Leo \citep{hawley1991}. 

For the SED models shown in \autoref{fig:lineplot}, the inclusion of emission lines results in an increase of 0.157 mag in $m_g$ and decrease of 44.2 $\rm\AA$ in $\lambda_\mathrm{eff}$ compared to the blackbody-only SED. Expressed as percent changes relative to the magnitude and $\lambda_\mathrm{eff}$ of the blackbody-only SED, the inclusion of lines changes $\lambda_\mathrm{eff}$ by 0.9\% and $m_g$ by 0.2\%. As the emission lines shift the $\lambda_\mathrm{eff}$ blue-ward and increase the brightness, the measured DCR is explained by smaller temperature. A smaller temperature for the black body, however, corresponds to a dimmer flare. The filling factor can compensate for the decreased blackbody brightness, and this is 
shown in \autoref{fig:result}. In the top row, the result of the model described in \autoref{sec:weffs} are shown as gray lines and the results of the model that includes the spectral features are shown in color. 

\subsection{Constraints on the Evolution of the Filling Factor $X_\mathrm{BB}$}\label{sec:constraints}

\begin{figure*}[!ht]
    \centering
    \includegraphics[width=0.85\textwidth]{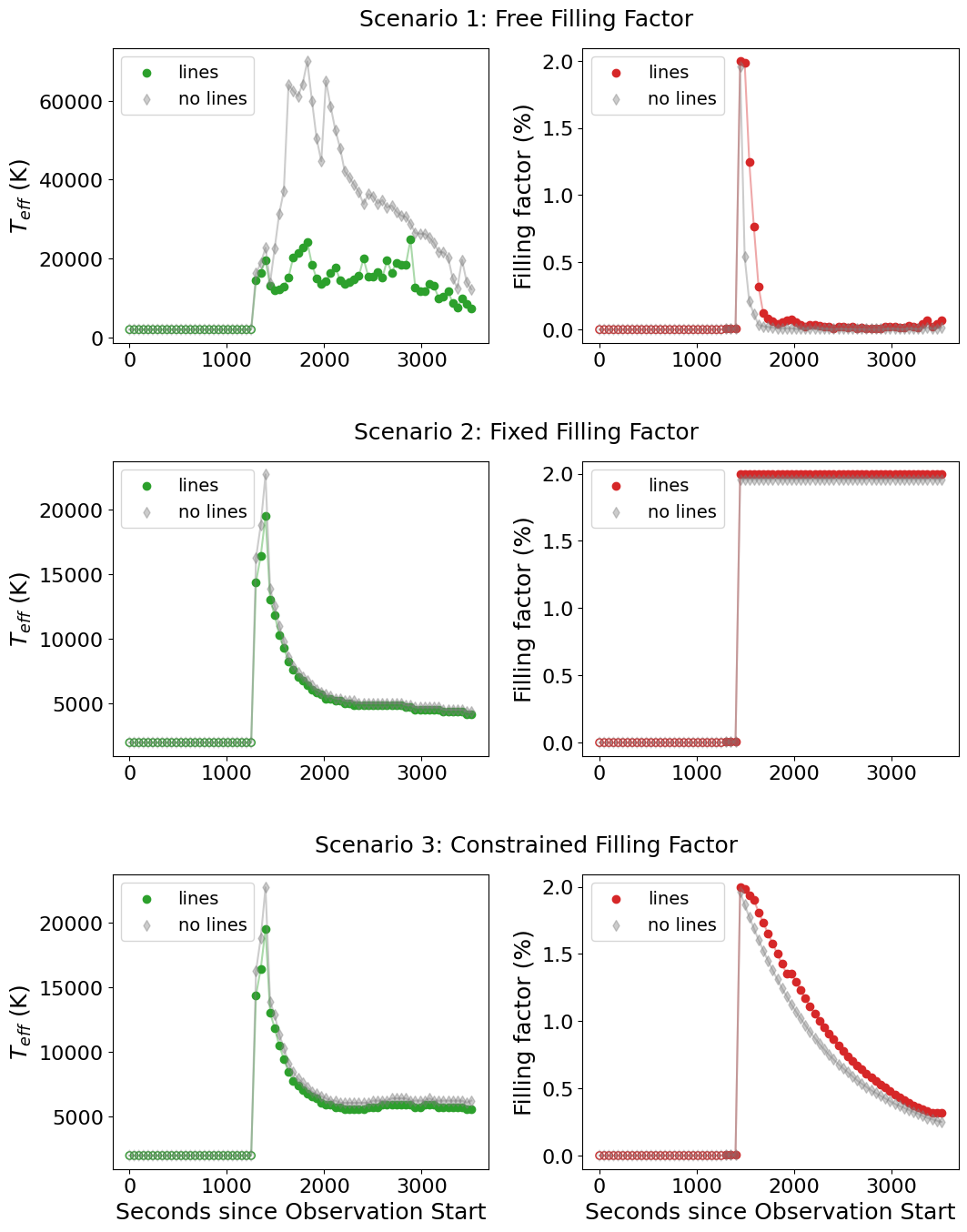}
    \caption{The derived evolution of $T_\mathrm{BB}$ and $X_\mathrm{BB}$ under models with and without the inclusion of emission lines (gray and colored points respectively) and which implement different post-peak constraints on $X_\mathrm{BB}$. In Scenario 1 (top), $X_\mathrm{BB}$ is allowed to vary freely throughout. In Scenario 2 (middle), the value of $X_\mathrm{BB}$ is held constant after peak. In Scenario 3 (bottom), the post-peak evolution of $X_\mathrm{BB}$  is constrained such that it cannot change by more than 0.107\% of its peak value per second.}
    \label{fig:result}
\end{figure*}

Since the size of the flaring region controls the blackbody contribution to the flare luminosity, and the magnitude and SED of the blackbody flare (\ie\ its temperature) jointly control $\lambda_\mathrm{eff}$, there is degeneracy in the $T_\mathrm{BB}$ - $X_\mathrm{BB}$ plane (\autoref{fig:grids}). Forward models that directly apply \autoref{eq:objective} lead to solutions where the filling factor rapidly drops after the flare peak (\autoref{fig:result1} and \autoref{fig:result} top row).

A study of the temporal and spatial evolution of an X2.0 class solar flare observed throughout ultraviolet ($1,600\rm\AA$) by the Transition Region and Coronal Explorer (TRACE) was presented in \citet{qiu2009}. When spatially resolved, the flare regions can be described as ``ribbons''. The authors measure the time evolution of three spatial properties of the flare ribbons: the projected length of the flare ribbons along the polarity inversion line (PIL), to which we will refer as $l_r$, the distances of the head and tail of the ribbons along the PIL relative to a fixed point, and the mean distance from the PIL to the ribbon front. They note that, while the size of the flare increases rapidly during the brightening phase,  the timescale of the ``shrinkage'' of the flare ribbon lengthwise along the PIL due to radiative cooling of the photosphere is significantly longer than the timescales of magnetic reconnection and heating of the stellar material, warranting constraints on the rate of change of $X_\mathrm{BB}$ in our forward model. 
\new{The temporal evolution of stellar flare geometric properties, such as $X_\mathrm{BB}$, is an area of active study and there remain open questions about the connection between stellar flare energetics and geometry. However, the solar-stellar analogy in flare physics is strongly reinforced by similarities in geometric phenomenology between the two paradigms \citep{reale2003, kowalski2013, kowalski2024} and solar flare observations can be used to place reasonable constraints on the evolution of $X_\mathrm{BB}$ in our forward model. For example, \citet{kowalski2013} measured the speeds at which the inferred blackbody flare area of a flare on YZ Canis Minoris grew during the rise, peak, and post-maximum phase, and found that these speeds are ``strikingly similar to the speeds of the development of two-ribbon flares parallel and perpendicular to the magnetic neutral line in solar active regions.'' We implement three models: 1) a fast evolving scenario arises naturally from the unconstrained original model, 2) we test a very slow evolution scenario by leaving $X_\mathrm{BB}$ unchanged for the duration of the flare and, 3)}
adapting the maximum rate of change observed in flare ribbon length along the PIL in \citet{qiu2009}, we restricted the post-peak evolution of $X_\mathrm{BB}$ to not exceed 0.107\% of it's value at peak magnitude, per second. \new{Thus we effectively span two orders of magnitude in $X_\mathrm{BB}$ evolution speed.}

In \autoref{fig:result}, we show 
the effective temperature and filling factor curves resulting from three different filling factor evolution constraints: (1) we allow the  $X_\mathrm{BB}$ to evolve freely; (2) we keep $X_\mathrm{BB}$ constant at its peak value (1.96\%); (3) following \citet{qiu2009}, we limit the rate of shrinkage of the flare by constraining the filling factor evolution speed to $dX_\mathrm{BB} <0.00107 X_\mathrm{BB}~s^{-1}$ after the flare reaches peak brightness. 

When $X_\mathrm{BB}$ is left to evolve without constraints (Scenario 1), its size rapidly decreases in a manner significantly inconsistent with the behavior observed in \citet{qiu2009}. 
In both Scenarios 2 and 3, when the evolution of $X_\mathrm{BB}$ is constrained, we observe the temperature evolves more rapidly and smoothly. As before, we see the effect of the inclusion of spectral lines reduces the inferred size of the filling factor, but also significantly the inferred temperature at peak. Where the spectral lines are included and the rate of $X_{BB}$ is constrained (Scenario 3), the flare reaches a maximum temperature of $19,510$~K, and spends a total of 2.4 minutes above $14,000$~K. In all scenarios in \autoref{fig:result}, we observe that the inclusion of emission features suppresses temperature while enhancing filling factor. This is because the emission features allow the model to reproduce the measured $\lambda_\mathrm{eff}$ with lower continuum temperatures, and in order to faithfully reproduce the photometry, the size of the emitting region must increase to compensate for the lower temperature.

\begin{figure*}[!ht]
    \centering
    \includegraphics[width=0.85\textwidth]{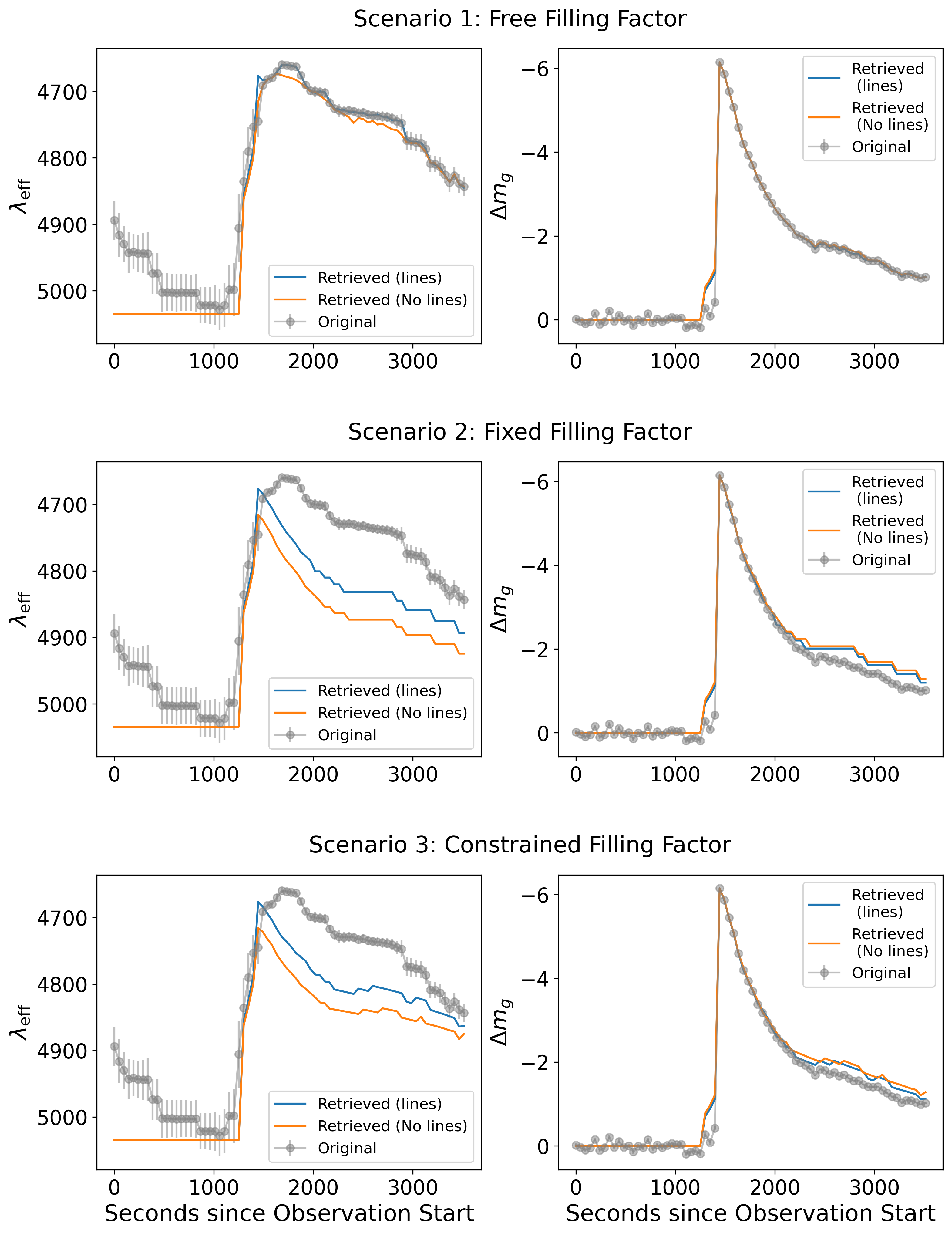}
    \caption{PSF photometry and effective wavelengths measured on the images (grey) compared to the forward model solutions for each of the three models described in \autoref{sec:constraints}, as well as for both SED models described in \autoref{sec:lines}. 
    }
    \label{fig:retrieval}
\end{figure*}

However, when we forward model $\lambda_\mathrm{eff}$ and $\Delta m_g$ (as described in \autoref{sec:forwardmodel}) from the inferred values of $T_\mathrm{BB}$ and $X_\mathrm{BB}$, and compare the derived and measured relative photometry and astrometry ($d_\parallel$ values and their evolution), we find that while without constraining the evolution of the filling factor the photometry and astrometry we observed are reproduced faithfully. When we set constraints, the model can reproduce the magnitude, but does no longer reproduce the effective wavelength we measured within the uncertainties. 
These comparisons are shown in \autoref{fig:retrieval}. Scenario 1, where the filling factor is not constrained, reproduces the observed photometry faithfully with values well within $1-\sigma$ uncertainty of the measurement at any time, and also reproduces the inferred $\lambda_\mathrm{eff}$ within $1-\sigma$ for the  emission line model. 
However, as noted, the rapidity of the filling factor evolution defies observations of flare ribbon evolution such as those of \citet{qiu2009}. Conversely, the solutions where the filling factor evolution is constrained (Scenarios 2 and 3) 
accurately reproduce magnitude evolution and $\lambda_\mathrm{eff}$ at peak, but tend to underestimate $\lambda_\mathrm{eff}$ after peak and are inconsistent with the post-peak observations, indicating that the constrained models' evolution struggles to reproduce the observations despite their consistency with known geometric evolution in solar flares. 

\section{Conclusions}\label{sec:conclusions} 

We have for the first time derived the temporal evolution of a stellar flare's color temperature via differential chromatic refraction. This was enabled by the flare's large magnitude excursion ($\Delta m_g = -6.12$), as well as unique combination of conditions in which the flare was observed, namely, in a blue filter (DES $g$, 398-548nm), and at high air mass ($1.45 \lesssim X \lesssim 1.75$), and by an instrument, DECam at the Blanco Telescope, that does not have an atmospheric dispersion corrector and also has sufficient spatial resolution and optical quality to enable sub-arcsecond position measurements. Using image data obtained from the Dark Energy Camera operating as part of the \textit{Deeper, Wider, Faster} program, we performed relative PSF photometry and astrometry throughout the observing period in which the flare was detected (\autoref{sec:astrometry}). We quantify the astrometric deviation of the flare star attributable to DCR by measuring the component of the motion in the direction of the zenith (\autoref{sec:analysis}). 

We use the $g$-band effective wavelengths inferred from the astrometry in combination with the PSF magnitude excursion to forward model the blackbody color temperature evolution of the flare under three different constraints on filling factor evolution and including a simple model for emission lines. To best match the observations, our model imposes a rapid evolution of the size of the flaring region after peak (the filling factor). However, we note that solar flare observations \citep{qiu2009} suggest the filling factor does not shrink rapidly during the flare decay. While the evolution of the filling factor remains highly uncertain, we suggest that a more sophisticated line evolution model may be the key to more faithfully reproducing the observed properties of the flare. 
For example, as already mentioned in \autoref{sec:statanalysis}, emission features in flare SEDs are known to vary in strength throughout a flare's lifetime. Indeed, the Ca II K line is notable for it's delayed onset relative to other emission features, peaking at later times in the flare evolution \citep{hawley1991, houdebine1991, houdebine2003, garcia2002, kowalski2013}. Adopting a time-dependent line strength profile in the forward modeling process could move towards a more physically rigorous approximation of the effective wavelength, and thus temperature derivation.

There remains room for refinement of this novel method to derive flare temperature evolution, as well as opportunities to compare these results with other flare temperature studies. For example, \citet{howard2020} found that superflare temperatures correlate with flare energy and impulse, the latter of which they quantify as flare amplitude divided by full-width-at-half-maximum, or the ``sharpness'' of the photometric rise. They additionally show that the amount of time spent above $14,000$~K tends to increase with flare energy. Comparisons of such parameters estimated via DCR versus other methods are another important step in validating results produced by DCR. 

This first measurement of flare temperature through DCR timely demonstrates the power of this method, as the upcoming Vera C. Rubin Observatory LSST \citep{ivezic2019} will offer an invaluable opportunity to derive, via the technique utilized here, statistical samples of flare temperature-magnitude observations in the Wide Fast Deep survey, as well as in depth studies of individual flares in the Deep Drilling Fields, as we demonstrated in \citet{clarke2024}.


\section*{Acknowledgments}

This material is based upon work supported by the National Science Foundation under Award No. AST-2308016.

J.R.A.D. acknowledges support from the DiRAC Institute in the Department of Astronomy at the University of Washington. The DiRAC Institute is supported through generous gifts from the Charles and Lisa Simonyi Fund for Arts and Sciences, and the Washington Research Foundation.

J.C. acknowledges funding from the Australian Research Council Centre of Excellence for Gravitational Wave Discovery (OzGrav) through project numbers CE170100004 and CE230100016 and funding by the Australian Research Council Discovery Project, DP200102102.

The authors acknowledge the support of the Vera C. Rubin Legacy Survey of Space and Time Science Collaborations and particularly of the Transient and Variable Star Science Collaboration (TVS SC), which provided opportunities for collaboration and exchange of ideas and knowledge.

The authors thank Dr. Michael Shay and Dr. Dana Longcope for sharing their knowledge of flares in the solar plasma which contributed towards a more complete interpretation of our data. 

This work has made use of data from the European Space Agency (ESA) mission
{\it Gaia} (\url{https://www.cosmos.esa.int/gaia}), processed by the {\it Gaia}
Data Processing and Analysis Consortium (DPAC,
\url{https://www.cosmos.esa.int/web/gaia/dpac/consortium}). Funding for the DPAC
has been provided by national institutions, in particular the institutions
participating in the {\it Gaia} Multilateral Agreement.

This project used data obtained with the Dark Energy Camera (DECam), which was constructed by the Dark Energy Survey (DES) collaboration. Funding for the DES Projects has been provided by the U.S. Department of Energy, the U.S. National Science Foundation, the Ministry of Science and Education of Spain, the Science and Technology Facilities Council of the United Kingdom, the Higher Education Funding Council for England, the National Center for Supercomputing Applications at the University of Illinois at Urbana Champaign, the Kavli Institute for Cosmological Physics, University of Chicago, the Center for Cosmology and Astroparticle Physics, Ohio State University, the Mitchell Institute for Fundamental Physics and Astronomy at Texas A\&M University, Financiadora de Estudos e Projetos, Fundacao Carlos Chagas Filho de Amparo a Pesquisa do Estado do Rio de Janeiro, Conselho Nacional de Desenvolvimento Cientfico e Tecnologico and the Ministerio da Ciencia, Tecnologia e Inovacao, the Deutsche Forschungsgemeinschaft, and the Collaborating Institutions in the Dark Energy Survey. The Collaborating Institutions are Argonne National Laboratory, the University of California, Santa Cruz, the University of Cambridge, Centro de Investigaciones Energeticas, Medioambientales y Tecnologicas-Madrid, the University of Chicago, and University College. 
This research made use of Photutils, an Astropy package for
detection and photometry of astronomical sources \citep{larry_bradley_2024_13989456}.


\bibliography{references}{}
\bibliographystyle{aasjournal}


\end{document}